\documentclass[12pt]{article}
\usepackage{mathtools,amssymb,amsthm,mathrsfs}
\usepackage{txfonts,amstext,amssymb,amsfonts,color,lscape,youngtab,tikz,bbold}
\usepackage[english, USenglish]{babel}
\usepackage{graphicx}

\advance\voffset by -1.5cm
\advance\hoffset by -1.25cm
\definecolor{darkgreen}{rgb}{0,0.6,0}
\def\thefootnote{\fnsymbol{footnote}}

\def\Tr{\,{\rm Tr}\, }

\def\be{\begin{equation}}
\def\ee{\end{equation}}
\def\ba{\begin{eqnarray}}
\def\ea{\end{eqnarray}}

\newcommand{\D}{{\cal D}}

\newcommand{\N}{{\cal N}}

\newcommand{\ZZ}{\mathbb{Z}}

\newcommand{\g}{\mathfrak{g}}

\newcommand{\SU}{{\rm SU}}

\newcommand{\U}{{\rm U}}

\renewcommand{\l}{\lambda}

\renewcommand{\SU}{{\rm SU}}

\newcommand{\bpsiup}{\overline{\psiup}}
\newcommand{\bpartial}{\overline{\partial}}
\newcommand{\barz}{{\overline{z}}}

\newtheorem{thm}{Theorem}[section]

\newtheorem{prop}[thm]{Proposition}

\usepackage[small,nohug,heads=vee]{diagrams}
\diagramstyle[labelstyle=\scriptstyle]
\def\N{\text{N}}
\def\D{\text{D}}

\newmuskip\pFqskip
\mathchardef\pFcomma=\mathcode`,

\newcommand{\secret}[1]{}

\begin{document}
\Yautoscale0
\addtolength{\baselineskip}{5pt}

%%%%%%%%%%%%%%%%%%%%%%   TITLE    %%%%%%%%%%%%%%%%%%%%

\thispagestyle{empty}
\renewcommand{\thefootnote}{\fnsymbol{footnote}}

{\hfill \parbox{2.45cm}{
 DESY 14-169 \\
}}

\bigskip

\begin{center} \noindent \Large \bf
Chiral Ring of Strange Metals: The Multicolor Limit
\end{center}

\bigskip\bigskip\bigskip

\centerline{ \normalsize \bf
 Mikhail Isachenkov$^a$
  \footnote[2]{\noindent \tt email: mikhail.isachenkov@desy.de},
 Ingo Kirsch$^a$
  \footnote[1]{\noindent \tt email: ingo.kirsch@desy.de}
and Volker Schomerus$^a$
  \footnote[1]{\noindent \tt email: volker.schomerus@desy.de}
}

\bigskip

\centerline{\it ${}^a$ DESY Hamburg, Theory Group,}
\centerline{\it Notkestrasse 85, D-22607 Hamburg, Germany}

\bigskip\bigskip

\bigskip\bigskip

\renewcommand{\thefootnote}{\arabic{footnote}}

\centerline{\bf \small Abstract}
\medskip
The low energy limit of a dense 2D adjoint QCD is described by a family of
${\cal N}=(2,2)$ supersymmetric coset conformal field theories. In previous
work we constructed chiral primaries for a small number $N < 6$
of colors. Our aim in the present note is to determine the chiral
ring in the multicolor limit where $N$ is sent to infinity. We shall
find that chiral primaries are labeled by partitions and identify the ring
they generate as the ring of Schur polynomials. Our findings impose
strong constraints on the possible dual description through string
theory in an $AdS_3$ compactification.

%%%%%%%%%%%%%%%%%%%%%%%%%%%%%%%%%%%%%%%%%%%%%%%
\newpage
\setcounter{tocdepth}{2}
%%%%%%%%%%%%%%%%%%%%%%%%%%%%%%%%%%%%%%%%%%%%%%%

\setcounter{equation}{0}
%%%%%%%%%%%%%%%%%%%%%%%%%%%%%%%%%%%%%%%%
\section{Introduction}
%%%%%%%%%%%%%%%%%%%%%%%%%%%%%%%%%%%%%%%%

Throughout the last few years, low dimensional examples of dualities
between conformal field theories and gravitational models in Anti-deSitter
(AdS) space have received quite some attention. There are at least two
motivations for such developments. On the one hand, many low dimensional
critical theories can actually be realized in condensed matter systems.
As these are often strongly coupled, the AdS/CFT correspondence might
provide intriguing new analytic tools to compute relevant physical
observables, much as it does for models of particle physics. On the
other hand, low dimensional incarnations of the AdS/CFT correspondence
might also offer new views on the very working of dualities between
conformal field theories and gravitational models in AdS backgrounds.
This applies in particular to the AdS$_3$/CFT$_2$ correspondence since
there exist many techniques to solve 2-dimensional models directly, without
the use of a dual gravitational theory. Recent examples in this direction
include the correspondence between certain two-dimensional coset conformal
field theories and higher spin gauge theories \cite{Gaberdiel:2010pz,
Gaberdiel:2011zw}, see also \cite{Creutzig,Creutzig2,Candu,Gaberdiel}
for  examples involving supersymmetric conformal field theories and
\cite{Gaberdiel:2014yla, Candu2} for a more extensive list of the vast literature
on the subject. It would clearly  be of significant interest to construct
new examples of the AdS$_3$/CFT$_2$ correspondence which involve full
string theories in AdS$_3$.

In 2012, Gopakumar, Hashimoto, Klebanov, Sachdev and Schoutens \cite{Gopakumar}
studied a two-dimensional adjoint QCD in which massive Dirac fermions $\Psi$
are coupled to an SU$(N)$ gauge field. The fermions were assumed to transform
in the adjoint rather than the fundamental representation of the gauge group.
In the strongly coupled high density region of the phase space, the corresponding
infrared fixed point is known to develop an \mbox{${\cal N}=(2,2)$} superconformal
symmetry. For a very small number $N \leq 3$ of colors, the fixed points belong to
the series of \mbox{${\cal N}=(2,2)$} superconformal minimal models and hence they
are very well studied. But in order to compare with tree level string theory, one
needs to explore the multicolor limit in which $N$ goes to infinity. This regime
is much less understood. Note that the central charge $c_N=(N^2-1)/3$ of fixed
point theories grows quadratically with the rank $N-1$ of the gauge group. While
this is very suggestive of a string theory dual, there exist very little further
clues on the appropriate choice of the 7-dimensional compactification manifold
$M^7$ of the relevant AdS background.

A central clue for discriminating between potential gravitational duals of the
infrared fixed point is expected to come from the chiral ring \cite{Lerche}, i.e.\ the algebra of operators $\phi$ satisfying the $BPS$ bound $h (\phi) = Q(\phi)$. Here $h$
denote the scaling weight and $Q$ the U$(1)$ R-charge of $\phi$, respectively. A
certain subset of so-called regular chiral primaries is easy to construct and
some of them had appeared in \cite{Gopakumar} already. But once we leave the territory
of minimal models, these do not exhaust the set of chiral primary operators. In
a previous paper we pushed the study of chiral primaries to $N > 3$ and
constructed all such operators for $N=4$ and $N=5$. In both cases, we found
new chiral primary operators that we dubbed {\em exceptional}. The total number
of such exceptional chiral primaries can be shown to grow very rapidly with $N$.\footnote{In particular, we observed that the number of exceptionals
grows faster than the number of regular chiral primaries whose number grows
as $2^N$. For example, while there is a single exceptional at $N=4$ along
with $7$ regulars, the $N=8$ theory possesses $153$ exceptional chiral
primaries which outnumber the $125$ regulars.}
On the other hand, the examples we reported on satisfy the BPS condition $h=Q$
only for one special value of $N$. Therefore it is not evident that exceptional
chiral primaries contribute to the chiral ring of the multicolor limit. This is
the question we are about to address with the present paper.

The main result of our analysis is that the large $N$ limit of the chiral
ring receives contributions only from regular chiral primaries. The latter
can be counted quite easily. As described in \cite{Isachenkov}, they are labeled
by partitions or Young diagrams. Moreover, their operator product expansions
may be argued to agree with the product of Schur polynomials. This provides
a complete description of the chiral ring in the large $N$ limit.

The plan of this short note is as follows. In the next section we shall introduce
the model and review some of the key results from \cite{Isachenkov}. Section 3 contains
the main new results of this paper. There we shall show that chiral primaries can
only contribute in the limit $N \rightarrow \infty$ if they are regular. The
operator products of regular chiral primaries are discussed in the concluding
section along with a few open problems that should be addressed in future studies
of the model.

\setcounter{equation}{0}
%%%%%%%%%%%%%%%%%%%%%%%%%%%%%%%%%%%%%%
%%%%%%%%%%%%%%%%%%%%%%%%%%%%%%%%%%%%%%
\section{Review of Background Material}
%%%%%%%%%%%%%%%%%%%%%%%%%%%%%%%%%%%%%%
%%%%%%%%%%%%%%%%%%%%%%%%%%%%%%%%%%%%%%

The role of this section is to review the definition of the model and the
construction of its state space. We shall also recall a few central
results on chiral primaries, including the construction of regular
chiral primaries, that have been discussed in \cite{Gopakumar} and
then extended in \cite{Isachenkov}.

\subsection{The coset model}

The model we start with is a 2-dimensional version of QCD with fermions
in the adjoint representations, i.e.\
\begin{equation}
 {\cal L}(\Psi,A) = \Tr \left[ \overline{\Psi}(i\gamma^\mu D_\mu -m -\mu
 \gamma^0)\Psi\right]
 - \frac{1}{2g_{\text{YM}}^2 } \Tr F_{\mu\nu} F^{\mu\nu} \ .
\end{equation}
Here, $A$ denotes an SU$(N)$ gauge field with field strength $F$ and gauge
coupling $g_\text{YM}$. The complex Dirac fermions $\Psi$ transform in the
adjoint of the gauge group and $D_\mu$ denote the associated covariant
derivatives. The two real parameters $m$ and $\mu$ describe the mass and
chemical potential of the fermions, respectively.

We are interested in the strongly coupled high density regime of the
theory, i.e.\ in the regime of very large chemical potential $\mu \gg
m$ and $g_\text{YM}$. As is well known, we can approximate the
excitations near the zero-dimensional Fermi surface by two sets of
relativistic fermions, one from each component of the Fermi surface.
These are described by the left- and right-moving components of
massless Dirac fermions. At strong gauge theory coupling, the
resulting (Euclidean) Lagrangian reads
\begin{equation}
{\cal L}_{\text{eff}}(\psiup,\bpsiup,A) = \Tr
\left( \bpsiup^\ast \partial\, \bpsiup + \psiup^\ast \bpartial\, \psiup
+ A_z [\,\psiup^\ast,\psiup\,] + A_\barz [\, \bpsiup^\ast,\bpsiup\,]\right)\ .
\end{equation}
Here we have dropped the term involving the field strength $F$, using
that $g_\text{YM}\rightarrow \infty$. Upon integrating out the two
components $A_z$ and $A_\barz$ of the gauge field we obtain the
constraints
\begin{equation} \label{eq:constraint}
J(z) \ := \ [\,\psiup^\ast,\psiup\,] \ \sim\ 0  \quad , \quad
\bar J(\bar z) \ := \ [\, \bpsiup^\ast,\bpsiup\,] \ \sim \ 0 \ .
\end{equation}
These constraints are to be implemented on the state space of the
$N^2-1$ components of the complex fermion $\psiup$ such that all the modes
$J_n, n > 0,$ of $J(z) = \sum J_n z^{-1-n}$ vanish on physical states, as
is familiar from the standard Goddard-Kent-Olive coset construction~\cite{GKO}.

In order to describe the chiral symmetry algebra of the resulting
conformal field theory we shall start with the unconstrained model
which we refer to as the {\em numerator} theory. It is based on
$M= N^2-1$ complex fermions $\psiup_\nu, \nu = 1, \dots, M$. These
give rise to a Virasoro algebra with central charge $c_\N = N^2-1$,
where the subscript $\N$ stands for numerator. We can decompose
each complex fermion into two real components $\psi^n_\nu, n=1,2,$
such that $\psiup_\nu = \psi^1_\nu + i \psi^2_\nu$. From time to
time we shall combine $\nu$ and $n$ into a single index $\alpha=
(\nu,n)$. Let us recall that the $2M$ real fermions $\psi_\alpha$
can be used to build SO($2M$) currents $K_{\alpha\beta}$ at level
$k=1$. The central charge of the associated Virasoro field coincides
with the central charge $c_\N$ of the original fermions. The
SO($2M$)$_1$ current algebra generated by the modes of $K_{\alpha\beta}$
forms the numerator in the coset construction.

The algebra generated by the constraints \eqref{eq:constraint} form the
denominator of the coset construction. According to the usual free fermion
constructions of current algebras, the fields $J$ that were introduced
in eq.\ \eqref{eq:constraint} form a SU($N$) current algebra at level
$k=2N$. The components $J_\nu = j^1_\nu + j^2_\nu$ can be written as a
sum of SU($N$) currents $j^n_\nu$ with $\nu=1,\dots,M$ and $n=1,2$. The
latter are obtained as bilinears of the real fermions $\psi^n_\nu, n=1,2,$
that we used in our description of the numerator theory. Through the
Sugawara construction we obtain a Virasoro algebra with central charge
$c_\D = 2(N^2-1)/3$, where the subscript $\D$ stands for denominator. Now
we have assembled all the elements that are needed in defining the coset
chiral algebra
\begin{equation}\label{eq:W2N}
{\cal W}_N := \text{SO}(2N^2-2)_1 / \text{SU}(N)_{2N}\ .
\end{equation}
The parameter $N$ keeps track of the gauge group SU($N$). The algebra
${\cal W}_N$  is a key element in our subsequent analysis. It is larger
than the chiral symmetry considered in \cite{Gopakumar} which uses the
subalgebra SU$(N)_N \times$ SU$(N)_N \subset$ SO$(2N^2-2)_1$ to encode
symmetries of the numerator theory.

Before we conclude this short review of the underlying model, let us
recall that the chiral algebra ${\cal W}_N$ contains a $\U(1)$ current.
It is constructed as
\begin{align} \label{eq:current}
J(z) =\textstyle \frac{1}{3} \sum_{\nu,\mu} \psi^1_\nu(z) \psi^2_\mu(z)
\kappa^{\nu\mu} \
\end{align}
where $\kappa^{\nu\mu}$ denotes the Killing form of $\SU(N)$.
The zero mode of this current turns out to measure the R-charge of
fields in ${\cal N} = (2,2)$ superconformal low energy limit of 2D
adjoint QCD. It will therefore play a very important role in the
subsequent analysis.

\subsection{The state space}\label{sec22}

Our second aim is to discuss the state space of the coset model. We
shall start by discussing the sectors of the chiral algebra ${\cal W}_N$
before concluding with a few comments on the modular invariant partition
function of the model. Since the Ramond (R) and Neveu-Schwarz (NS) sector of
an ${\cal N} =(2,2)$ superconformal field theory are related by spectral
flow \cite{Lerche}, our discussion will focus on the NS sector.

Let us denote the state space that is created with chiral fields of the
numerator theory in the NS sector by $\mathcal{H}^\text{NS}$. Under the
action of the denominator chiral algebra $\SU(N)_{2N}$ the space
$\mathcal{H}^\text{NS}$ decomposes as
\begin{equation}\label{decomp}
\mathcal{H}^\text{NS} \ \cong \ \bigoplus_{a\in \mathcal{J}_N} \, \mathcal{H}^\text{C}_{\{a\}} \otimes
\mathcal{H}_a^\text{D} \ .
\end{equation}
Here, $\mathcal{H}_a^\text{D}$ denotes the sectors of the denominator algebra
$\SU(N)_{2N}$ and $a \in {\cal J}_N$ is the corresponding weight. We shall consider
${\cal J}_N$ as the set of $N-1$ tuples
\begin{equation} \label{a}
 a = [\lambda_1,\dots, \lambda_{N-1}] \quad \mbox{with } \quad
\sum_s^{N-1} \lambda_s \leq 2N\ .
\end{equation}
Alternatively, the elements of $\mathcal{J}_N$ may be thought of as $\SU(N)$ Young
diagrams $Y = Y_a$. Given $a=[\lambda_1,\dots,\lambda_{N-1}]$ the length of the $i$th
row is
\begin{equation} \label{Y}
Y_a = (l_1,\dots,l_{N-1}) \quad  \mbox{is}\quad l_i = \sum_{s = i}^{N-1}\lambda_s \ .
\end{equation}
Of course it is just as easy to reconstruct $a=a\,(Y)$ from a Young diagram $Y$.
The factor $\mathcal{H}^\text{C}_{\{a\}}$ has been introduced to denote sectors of
the coset chiral algebra ${\cal W}_N$. It will become clear momentarily why we placed
the index $a$ in brackets $\{ \cdot \}$.

As usual in the coset construction, for $\mathcal{H}^\text{C}_{\{a\}}$ not to be
empty, the label $a$ must satisfy certain selection rules. In addition, some of
the spaces $\mathcal{H}^\text{C}_{\{a\}}$ carry equivalent representations of
${\cal W}_N$. In order to describe the relevant selection rules and field
identifications, we need to introduce the following map $\gamma$
\begin{align} \label{eq:SC}
\gamma([\lambda_1,\dots,\lambda_{N-1}]) = [2N-\sum_{s=1}^{N-1}\lambda_s,
\lambda_1,\dots, \lambda_{N-2}] \ .
\end{align}
Obviously, $\gamma$ maps elements $a \in {\cal J}_N$ back into ${\cal J}_N$ and it
obeys $\gamma^{N}={\it id}$. One can show that two sectors ${\cal H}^\text{C}_{\{a\}}$
and ${\cal H}^\text{C}_{\{b\}}$ of the coset chiral algebra are isomorphic provided that
the weights $a$ and $b$ are related to each other by repeated application of $\gamma$
or, equivalently,
\begin{equation} \label{identification}
{\cal H}^\text{C}_{\{a\}} \cong {\cal H}^\text{C}_{\{\gamma(a)\}} \quad
\mbox{for} \quad a \in {\cal J}_N \ .
\end{equation}
The isomorphism respects the action of the coset chiral algebra ${\cal W}_N$
on the sectors ${\cal H}^\text{C}_{\{a\}}$. In order to state the selection rules we
recall that the conformal weight $h^\D: {\cal J}_N \rightarrow \mathbb{R}$
is given by
\begin{equation}
h^\D(a) =   \frac{C_2(a)}{3N}\,,
\end{equation}
where the quadratic Casimir of an $\SU(N)$ representation $a =
[\lambda_1,\dots,\lambda_{N-1}]$ takes the form
\begin{align}
C_2(a)
&=\frac{1}{2} \left[-\frac{n^2}{N} + n N  +
\sum_{i=1}^r (  l^2_i + l_i - 2i l_i) \right] \,. \nonumber
\end{align}
Here, we have used the row length parameters $l_i$ introduced in eq.\
\eqref{Y} and $n$ denotes the total number of boxes $|Y|=n=\sum_{i} l_i$
in the Young diagram $Y = Y_a$. With these notations let us introduce the
so-called monodromy charge
\begin{align}\label{monocharge}
Q_\gamma(a) \equiv h^\D(\gamma(a))-h^\D(a) \ \mbox{mod} \ 1 \ .
\end{align}
For the carrier space  ${\cal{H}}^\text{C}_{\{a\}}$ of the coset algebra to be
non-vanishing, the label $a$ should be taken from the set ${\cal J}^0_N$ of
$\SU(N)_{2N}$ labels $a$ with vanishing monodromy charge $Q_\gamma(a)= 0$,
\begin{equation}\label{selection}
{\cal H}^\text{C}_{\{a\}} \cong \emptyset \quad \mbox{if} \quad Q_\gamma(a)
\neq 0\ .
\end{equation}
Since representations of the coset chiral algebra ${\cal W}_N$  are invariant
under the action  \eqref{eq:SC} of the identification group $\mathbb{Z}_{N}$,
isomorphism classes of representations of the coset chiral algebra are labeled
by orbits $\{a\} \in {\cal O}_N = {\cal J}_N^0/\ZZ_N$.
\medskip

A very useful way to parametrize elements of ${\cal J}_N^0$, i.e.\ $\SU(N)_{2N}$
weights $a$ with vanishing monodromy charge, through a pair of Young diagrams $Y'$
and $Y''$ was described in \cite{Isachenkov}. Following that approach, we introduce
two SU$(N)$ Young diagrams $Y'$ and $Y''$ with equal number $n' = |Y'|= |Y''|$
of boxes, subject to the additional conditions
\begin{equation} \label{condgeneration}
r' + c'' \leq N \quad ,\quad r'' + c' \leq 2N \
\end{equation}
where $r',r''$ and $c',c''$ denote the numbers of rows and columns
of $Y',Y''$, respectively. Let us denote the row lengthes of the
Young diagrams $Y'$ and $Y''$ by
$$ Y' = (l'_1,\dots,l'_{r'}) \quad , \quad  Y'' = (l''_1,\dots,l''_{r''})
\ .$$
As before, we arrange the $l'_i$ and $l''_i$ in decreasing order, i.e.\
$l'_i \geq l'_{i+1}$ etc. so that the largest entries are $l_1'  =
c'$ and $l_1'' = c''$. From these two Young diagrams we can build
a new diagram $Y=Y(Y',Y'') = (l_1, \dots, l_{N-1})$ through
\emph{}\begin{align}
l_i = \left\{
\begin{array}{lll}
r'' +l'_i  &\rm for& i=1, ..., r'\\[2mm]
r'' &\rm for & i=r' +1, ..., N-l''_1\\[2mm]
r'' -k &\rm for & i=N-l''_k+1, ..., N-l''_{k+1} \,, \quad k=1, ..., r'' -1 \\[2mm]
0 &\rm for & i=N-l''_{r''}+1, ..., N-1\,.
\end{array}
\right. \label{generation}
\end{align}
This prescription extends a construction in \cite{Schwimmer} and it gives
a special family of so-called composite representations $Y=\bar Y'' Y'$ in 
the sense of \cite{Gross}. The latter have been defined without the additional 
condition $|Y'| = |Y''|$. It is not too difficult to show that all diagrams 
$Y = Y(Y',Y'')$ obtained in this way correspond to an $\SU(N)_{2N}$ weight 
$a = a\,(Y) = a\,(Y',Y'')$ with vanishing monodromy charge. Conversely, any 
such weight arises from a suitably chosen pair $(Y',Y'')$.

The inverse procedure of obtaining diagrams $Y'$ and $Y''$ from a given diagram $Y=Y_a=(l_1, \dots, l_{N-1})$ satisfying the zero monodromy charge condition
\begin{align}
\sum_i l_i\equiv 0 \text{ mod } N,
\end{align}
goes as follows. One defines $r'':=\frac{1}{N}\sum_i l_i$, $c':=l_1-r''$. Then the entries of the small Young 
diagrams $Y'$ and $Y''$ can be written as
\begin{align}
\left\{
\begin{array}{lll}
Y':=(l_1-r'',l_2-r'',\dots,l_{N-1}-r''),&&  \\[2mm]
Y''^{\,\text{T}}:=(r'',r''-l_{N-1},\dots,r''-l_{1}), &&  \\[2mm]
\end{array}
\right. \label{inverse_gen}
\end{align}
In both expressions the order of entries is non-decreasing and all non-positive
entries are to be skipped from the end of these strings.
\begin{figure}
\begin{center}
\includegraphics[scale=0.7]{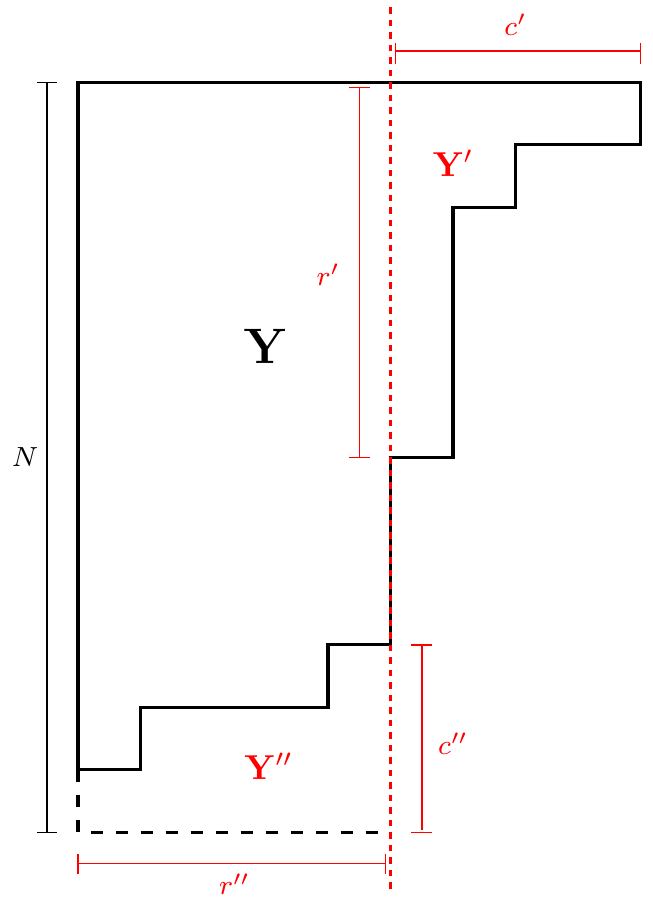}
\caption{Dissecting the Young diagram $Y$ by the red dashed line identifies Young diagrams $Y'$ and $Y''$ as soon as their numbers of boxes match. This can happen exactly once.
}\label{fig1}
\end{center}
\end{figure}

The prescriptions \eqref{generation} and \eqref{inverse_gen} might appear somewhat
heavy at first, but they possess a very simple pictorial representation,
see Figure \ref{fig1}. Suppose we are given the two Young diagrams $Y'$ and $Y''$.
Then we need to flip $Y''$ and place it on the bottom line of the image
which is $N$ boxes below the top line. This Young diagram has to start in
the first column and hence will extend over $r''$ columns. We now fill all
the boxes above the flipped diagram before we attach the second Young
diagram $Y'$ on the right hand side. Conversely, if we are given $Y$, we
must first construct the flipped $Y''$. It is made from all the boxes that
are needed to fill the space below the Young diagram $Y$, including the
$N$-th row. On the right hand side, we include as many columns $r''$ as are
needed for the flipped $Y''$ to possess as many boxes as the Young diagram
$Y'$ that appears to the right of the $r''$th column. This can be done by
increasing the number of columns one by one until the appropriate $r''$ is
found. If no appropriate choice of $r''$ exists, the original Young diagram
$Y$ does not correspond to a sector with vanishing monodromy charge.

With the map $Y(Y',Y'')$ and its inverse well understood, we want to
mention two properties of $Y$ that become relevant later on. To begin with,
it is evident from the geometric construction we described in the previous
paragraph that the diagram $Y= Y(Y',Y'')$ possesses $|Y|=n= r'' N$ boxes.
Furthermore, one can show \cite{Isachenkov} that the value of the quadratic
Casimir in the corresponding representation of $\SU(N)$ is given by
\begin{align}
C_2(a(Y',Y'')) = C_2(Y(Y',Y'')) = n' N + C_2(Y') - C_2(Y'') \,,
\label{C2generalized}
\end{align}
where $C_2(Y')$ and $C_2(Y'')$ are the quadratic Casimirs of the
$\SU(N)$ representations associated with $Y'$ and $Y''$, respectively.
\medskip

At this point, we have explained everything there is to know about the
formula \eqref{decomp}. Even though we wrote that the summation index $a$
is taken out of ${\mathcal{J}}_N$, we should keep in mind that the summands are
trivial unless $a \in \mathcal{J}_N^0$ simply because the corresponding spaces
$\mathcal{H}^\text{C}_{\{a\}}$ vanish. Hence, we can think of the summation as
running over pairs $(Y',Y'')$ of Young diagrams subject to the conditions
\eqref{condgeneration}. Finally, while the representation spaces
$\mathcal{H}^\text{D}_a$ of the denominator current algebra depend on the
weight $a = a(Y',Y'')$, the sectors $\mathcal{H}^\text{C}_{\{a\}}$ are
invariant under the action \eqref{eq:SC} of $\gamma$ and hence only depend
on the $\mathbb{Z}_{N}$ orbit $\{a\} = \{a(Y',Y'')\}$ of $a$.

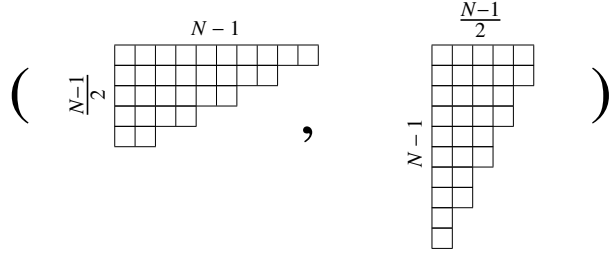
\begin{figure}
\begin{picture}(60,110)

 \put(108,63) { \Huge  ( }

 \put (134,58) {\rotatebox{90}{\begin{small} $\frac{N-1}{2}$\end{small}}}
 \put (178,92) {\begin{scriptsize} $N-1$\end{scriptsize}}

 \put(148,50)  {\Yboxdim8pt \Yvcentermath1 \yng(10,8,6,4,2)}
 \put(218,55) { \Huge , }

 \put (262,40) {\rotatebox{90}{\begin{scriptsize} $N-1$ \end{scriptsize}}}
 \put (278,95) {\begin{small} $\frac{N-1}{2}$ \end{small}}

 \put(268,11.5)  {\Yboxdim8pt \Yvcentermath1 \yng(5,5,4,4,3,3,2,2,1,1) }

 \put(328,63) { \Huge ) }
\end{picture}

\caption{
The pair $\left(Y'(a_\ast), Y''(a_\ast)\right)$ corresponding to 
$a_\ast=[2,2,\dots, 2]$, $N$ assumed odd.}
\label{fig2}
\end{figure}

The decomposition \eqref{decomp} is just used to build the representations
${\cal H}^\text{C}_{\{a\}}$  of the coset chiral algebra ${\cal W}_N$ but it
does not tell us yet how these sectors are combined with those of the right
moving chiral algebra in order to build a fully consistent conformal field
theory. The relevant modular invariant partition function was described in
\cite{Isachenkov}. For our purposes, it suffices to consider the case when
$N$ is prime. With this assumption, all but one of the sectors
$\mathcal{H}^\text{C}_{\{a\}}$ carry irreducible representations of the chiral
algebra. Only the sector $\mathcal{H}^\text{C}_{\{a_\ast\}}$ for $a_\ast =
[2,2,\dots, 2]$ can be decomposed into several irreducibles
$\mathcal{H}^\text{C}_{\{a_\ast\};\nu}$. The range of the index $\nu$ and
other features of this decomposition are described in \cite{Isachenkov}.
The full state space of the conformal field theory has been argued to
take the form
\begin{equation}
\mathscr{H}^\text{C} = \frac{1}{N}\ {\bigoplus_{Y',Y''}}'
\mathcal{H}^\text{C}_{\{a(Y',Y'')\}} \otimes
\overline{\mathcal{H}}\ \!^\text{C}_{\{a(Y',Y'')\}}
\ \oplus \ \mathscr{H}^\text{C}_\text{fix} \ .
\label{state_space}
\end{equation}
Here, we sum over all pairs $(Y',Y'')$ of Young diagrams that obey the
conditions \eqref{condgeneration} with the exception of the unique pair
that gives the Young diagram $Y = Y_{a_\ast}$, see Figure \ref{fig2}.
  Because of the identification rule \eqref{identification}, the sum over $Y'$
and $Y''$ gives each term with multiplicity $N$. This degeneracy is
removed when we divide by $N$. The term $\mathscr{H}^\text{C}_\text{fix}$
is built out of the sectors $\mathcal{H}^\text{C}_{\{a_\ast\};\nu}$ and their
right moving counterparts. The precise form, which can be found in
\cite{Isachenkov}, will not be relevant in the subsequent analysis.
Indeed, as we shall argue, this sector of the state space cannot contribute
any chiral primaries to the large $N$ limit.

%%%%%%%%%%%%%%%%%%%%%%%%%%%%%%%%%%%%%%%%
\subsection{Regular chiral primaries}\label{secregular}
%%%%%%%%%%%%%%%%%%%%%%%%%%%%%%%%%%%%%%%%

As we explained in the introduction, there exist two different classes of
chiral primary fields which we referred to as regular and exceptional.
While there is no general construction of the exceptional ones so far, the
regular chiral primaries may be listed explicitly for any value of $N$.
In fact, after we have introduced our parametrization of elements in
${\cal J}_N^0$ through pairs $(Y',Y'')$ of Young diagrams, this task
is really easy. In turns out that all sectors $\{a(Y',Y''=Y')\}$
of the coset chiral algebra contain precisely one (regular) chiral
primary field $\phi_\text{cp}\{a\}$. This field is to be found among
the ground states of the sector.

For ground states in the coset sector $\mathcal{H}^\text{C}_{\{a(Y',Y')\}}$
there exists a simple formula to compute the exact conformal weight. By eq.\
(\ref{C2generalized}) the quadratic Casimir of a representative $a=a(Y',Y')$
is simply $C_2(a)=n' N$ with $n'=|Y'|=|Y''|$. This implies that the conformal
weight of the ground states takes the form
\begin{align}
h(\phi{\{a\}}) &=
 \frac{C_2(a)}{6N} = \frac{n'}{6} \,,\label{cw}
\end{align}
for $a = a(Y',Y')$. Since the diagonal sectors $a\,(Y',Y')$ contain a chiral
primary ground state, its conformal weight and $\U(1)$ charge are given by
$n'/6$. Let us stress once again that the diagonal or regular sectors do
not contain any further chiral primaries among the $\mathcal{W}_N$
descendents~\cite{Isachenkov}.

Before we conclude this section we need to add a few comments that will later
allow us to enumerate regular chiral primaries, i.e.\ the orbits of diagonal
sectors $a(Y',Y')$. We should stress that most elements in such an orbit
$\{a(Y',Y''=Y')\}$ are not obtained from diagonal pairs $(Y',Y'') = (Y',Y')$.
So, if we would like to decide whether the sector $\mathcal{H}^\text{C}_{\{b\}}$
contains a chiral primary field, we need to construct the pair $(Y'(a),Y''(a))$
for each element $a$ in the orbit $\{b\}$ of the element $b\in {\cal J}^0_N$ and
check whether at least one of these pairs satisfies the condition $Y'(a) = Y''(a)$.
Let us note that the orbit $\{a_\ast\}$ of the weight $a_\ast = [2,2,\dots,2]$
consists of a single element $a_\ast$ and $Y'(a_\ast) \neq Y''(a_\ast)$ for any
prime $N > 2$. Hence, this special orbit does not contain a regular chiral primary
unless $N=2$.

More importantly, one can show that most orbits $\{b\}$ contain at
most one representative $a \in \{b\}$ such that $Y'(a) = Y''(a)$.
This follows from the following expression for the action of $\gamma^k$
in the weights of $(Y',Y'')$,
\begin{align*}
\begin{array}{lll}
\gamma^k(Y')&:=(2(N-k)+l_{N-k+1}-r'',\dots,2(N-k)+l_{N-k+1}-r'', \\[2mm] &
\hspace*{2cm} 2(N-k)-r'',l_1-2k-r'',\dots, l_{N-k-1}-2k-r'') \\[2mm]
\gamma^k(Y''^{\,\text{T}})&:=(r''+2k-l_{N-k},\dots,r''+2k-l_1,r''-2(N-k), \\[2mm]
& \hspace*{2cm} r''-2(N-k)-l_{N-1},\dots ,r''-2(N-k)-l_{N-k+2}).
\end{array}
\end{align*}
In both expressions the order of entries is non-decreasing and all non-positive
entries are to be skipped from the end of these strings. The
$Y^\text{T}$ is used to denote the transpose of a Young diagram $Y$. If we now
require $Y'=Y''$ and $\gamma^k(Y')=\gamma^k(Y'')$ it is easy to infer that the
only solutions satisfying these two constraints are the Young diagrams of
rectangular shape. These correspond to the orbits $\{a_\nu\}$ of the weight
$a_\nu = [0,...,0,N,0,...,0]$ for $\nu = 1, \dots, N-1$ where the only
non-zero entry $N$ can appear in any position $\nu$, i.e.\ $\lambda_\nu = N$.
As we have just demonstrated, field identifications can map $a_\nu$ to $a_{N-\nu}$.
Both of these weights are associated with diagonal pairs $(Y',Y''=Y')$ of Young
diagrams.

In conclusion we have argued that regular chiral primary fields of the coset
conformal field theory are associated with Young diagrams $Y'$ such that
$r'+c' \leq N$. The correspondence is one-to-one with the exception of
the Young diagrams $Y'_\nu$ and $Y'_{N-\nu}$ which correspond to one and
the same regular chiral primary.

\setcounter{equation}{0}
\section{Chiral primaries at large $N$}

We now address the central goal of this work, namely to construct the
chiral ring in the limit of large $N$. As we are about to vary $N$,
many of the objects we encountered in the previous section will carry
an additional label $N$. This applies in particular to the quadratic
Casimir $C^{(N)}_2$, the sectors $\mathcal{H}^{\text{C},(N)}$ of the
coset chiral algebra as well as the maps $a_N = a_N(Y',Y'')$ and $Y_N
= Y_N(Y',Y'')$ that associate a weight $a$ or a Young diagram $Y$  to
a pair of Young diagrams $Y'$ and $Y''$.

Since the coset model is built from representations of the coset chiral algebra,
we should first explain how to take the large $N$ limit of the sectors
${\mathcal H}^{\text{C},(N)}_{\{a\}}$. In the previous section we learned how
to parametrize the allowed values of $a$ in terms of two Young diagrams $Y'$
and $Y''$. In taking the limit, we keep these Young diagrams fixed, i.e.\ we
define
$$ \mathcal{H}_{\{Y',Y''\}} \equiv
\lim_{N \rightarrow \infty} \mathcal{H}^{\text{C},(N)}_{\{a_N(Y',Y'')\}}\ . $$
Let us stress that the Young diagram $Y=Y_a$ that we construct from $Y'$
and $Y''$ depends on the value of $N$. This is why it was so important to
place a subscript $\ _N$ on the corresponding $\SU(N)$ weight $a = a_N$.
One can show that the sectors $\mathcal{H}_{\{Y',Y''\}}$ are well defined. In
particular, the dimension of the subspaces with fixed conformal weight $h$
stabilizes as we send $N$ to infinity. We are now trying to find those
pairs $(Y',Y'')$ for which the space $\mathcal{H}_{\{Y',Y''\}}$ contains
chiral primaries. Our  claim is that this happens if and only if $Y'=Y''$.
As  we reviewed in the previous subsection, such diagonal pairs of Young
diagrams are associated with regular chiral primaries.

In order to establish these claims let us consider any of the summands
$$\mathcal{H}^\text{NS}_a = \mathcal{H}^{\text{C},(N)}_{\{a\}} \otimes \mathcal{H}^\D_a $$
in the decomposition \eqref{decomp}. The space ${\mathcal H}_{a}^\text{NS}$ comes
equipped with the action of several commuting operators. To begin with, we mention
the zero modes of the coset Virasoro field and the U(1) currents, i.e.\ $L_0 = L^G_0
- L^H_0$ and $Q$. In addition, we can also introduce the fermion number operator
$K_0$ which is defined by
$$ K_0 = \sum_{r \geq 1/2} \psi^1_{\mu,-r} \psi^1_{\nu,r} \kappa^{\mu\nu} +
     \psi^2_{\mu,-r} \psi^2_{\nu,r}\kappa^{\mu\nu} \ .$$
$K_0$ commutes with $Q$ and $L_0$  and hence can be measured simultaneously on
${\mathcal H}^\text{NS}_a$.

\begin{prop} \label{prop1}
The conformal weight $h_\phi$ of states $\phi$ in the subspace ${\mathcal H}_{a}^\text{NS}$
of the NS-sector is bounded from below by \begin{equation}
 h_\phi \geq \frac{K_\phi}{2} - \frac{C^{(N)}_2(a)}{3N}\ .
\end{equation}
Similarly, the U(1) charge $Q_\phi$ of the state $\phi$ is bounded from
above by
\begin{equation}
 |Q| \leq  \frac{K_\phi}{6}\ .
\end{equation}
In both inequalities, the number $K_\phi$ denotes the fermion number, i.e.\ the
eigenvalue of the fermion number operator $K_0$ on the state $\phi$.
\end{prop}

The two inequalities follow straightforwardly from the fact that the complex
fermion multiplets $\Psi$ and $\Psi^\ast$ have conformal weight $h_\Psi=1/2$
and that their real and imaginary part $\psi^1$ and $\psi^2$ possess U(1) charge
$|Q_{\psi^j}| = 1/6$. In the first relation, the two sides are equal in case
the construction of $\phi$ does not involve any derivatives of the fermionic
fields. The second relation becomes an equality for states $\phi$ that are
built from $\psi^1$ or $\psi^2$ and its derivatives only.

There is another simple proposition we need to discuss. Before we state it,
let us recall from \cite{Isachenkov} that a sector
$\mathcal{H}^{\text{C},(N)}_{\{a\}}$ of the coset model can only contain a
chiral primary if
$$ \text{min}_{b \in \{a\}} \left( C^{(N)}_2(b)\right) \equiv 0 \
\text{mod } N \  , $$
i.e.\ the minimum $C_2(b)$ assumed in the orbit $\{a\}$ of $a$ must be
divisible by $N$, at least when $N$ is odd. Under the action of the
identification current, the value of the quadratic Casimir can only shift
by an integer multiple\footnote{The precise amount of this shift is $C_2(\gamma^i(a))-C_2(a)=3N\left(\sum_{j=1}^{i-1}l_{N-j}+i\,(N-r''-i)\right)$.} of $N$  so that a sector $\mathcal{H}^{(N)}_{\{a\}}$
can only contain a chiral primary if
$$ C^{(N)}_2(a) \equiv 0 \ \text{mod } N \  . $$
As we explained before, when we vary $N$ we are instructed to keep
$Y'$ and $Y''$ fixed. Let us assume that $N_0$ is the minimal number
for which the two inequalities
$$ r'+c'' \leq N_0 \quad , \quad r'' + c' \leq 2N_0  $$
are satisfied. Then $Y'$ and $Y''$ define a sector of the coset
theory for all $N \geq N_0$. We can use the rules stated above to
construct a diagram $Y_N(Y',Y'')$ for all $N \geq N_0$. The
associated representation is denoted by $a_N = a_N(Y',Y'')$,
as before.

\begin{prop}\label{prop2}
The family of sectors $\mathcal{H}^{\text{C},(N)}_{\{a_N(Y',Y'')\}}$ can only contain
a chiral primary if the two Young diagrams $Y'$ and $Y''$ of SU$(N_0)$ belong
to representations with the same value of the quadratic Casimir element,
$$ C^{(N_0)}_2(Y') = C^{(N_0)}_2(Y'')\ .  $$
\end{prop}

To prove this statement we recall from  eq.\ \eqref{C2generalized} that the value
of the quadratic Casimir element in the representation $a_N$ of SU$(N)$ is given by
$$ C^{(N)}_2(a_N(Y',Y'')) = n' N  + C^{(N)}_2(Y')- C^{(N)}_2(Y'') \ .$$
For the sector $a_N$ to contain a chiral primary, the right hand side must be
divisible by $N$. Since the first term is, we need to determine the conditions
under which the $C^{(N)}_2(Y')-C^{(N)}_2(Y'')$ is a multiple of $N$.
The difference of the Casimir reads
\begin{align}
C^{(N)}_2(Y') - C^{(N)}_2(Y'') &= \frac{1}{2} \left( \sum_i (l_i^{\prime 2}+l_i^{\prime} -
2i l_i^{\prime}) - (l_i^{\prime\prime 2}+l_i^{\prime\prime} - 2il_i^{\prime\prime}) \right)
= C^{(N_0)}_2(Y') - C^{(N_0)}_2(Y'') \ ,
\end{align}
and thus does  not depend
on $N$. Hence it clearly cannot be divisible by (a sufficiently large)\footnote{A rough
estimate for what `sufficiently large' means is given by $N > N_0(N_0^2-1)/12$.} $N$
unless the difference vanishes. This is what we had to prove.

Our third proposition is a little more difficult to prove, but it
is absolutely crucial for what we are about to establish.

\begin{prop} \label{prop3}
For the states $\phi$ in the sector  ${\mathcal H}_{a}^\text{NS}$, the
fermion number satisfies the inequality
$$K_\phi \geq n'\ . $$
The number $n'$ is determined by the choice of $a$. It is computed
from the associated Young diagram $Y = Y_a$ by, see eq.\ (\ref{inverse_gen}),
$$n'=
\sum_{i=1}^{N-1} \theta \left(l_i-\frac{1}{N}\sum_{i=1}^{N-1}l_i\right),$$ where $\theta(x)$ denotes the Heaviside step-function. If the representation $a$ is
diagonal, i.e.\ $Y'(a)=Y''(a)$, the above formula simplifies to
$$n'=\sum_{i=1}^{n/N} \left(l_i-\frac{n}{N}\right)\ ,\quad  \text{ where } \quad n=\sum_{i=1}^{N-1}l_i\ .$$
\end{prop}

We will first give a somewhat heuristic graphical argument
using Young diagrams before we outline a formal proof of this proposition.
Let us recall that all our fermions transform in the tensor product of the
fundamental and the dual fundamental representations. These correspond to
Young diagrams that consist of a single box and a single column of maximal
length $N-1$, respectively.

We need to show that it takes at least $n'$ fermionic fields in
order to build a state in the representation $a$, i.e.\ the first time
the representation $a$ appears in the tensor power $ \textbf{adj}^{\otimes K}$
is for $K = n'$. For the SU$(N)$ Lie algebra, the adjoint representation decomposes as
$\textbf{adj}=\Box \otimes \overline \Box$. Here~$\overline
\Box$ denotes the (Young diagram of) the dual fundamental representation,
i.e.\ a column of $N-1$ boxes.

\begin{figure}
\begin{center}
\includegraphics[scale=0.7]{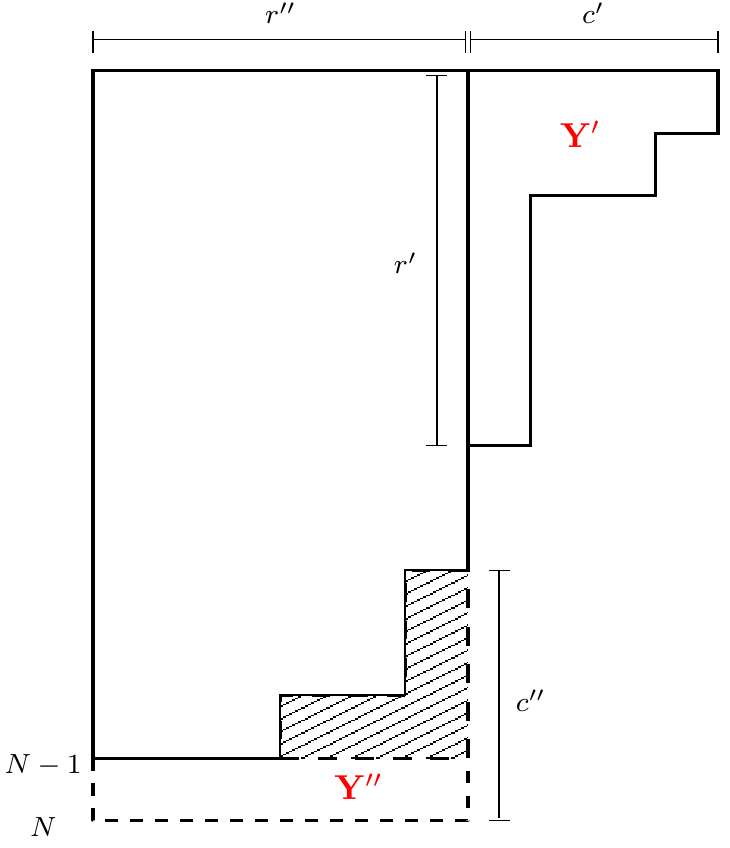}
\caption{The Young diagram $Y$ is obtained from a rectangle of size
$N \times r''$ by attaching the Young diagram $Y'$ and removing (a reflected
version of) the Young diagram $Y''$. Note that $r'+c'' \leq N$ is needed for
the resulting diagram to be a Young diagram of SU$(N)$. The shaded region
shows those boxes from the tensor power of the dual fundamental that must
be moved to the left.
}\label{fig3}
\end{center}
\end{figure}

The graphical proof goes as follows. In order to build the Young diagram
$Y$ we start with the $K$ columns $\overline \Box$ of size $N-1$. If we
put them all side by side, we would obtain a  rectangular
Young diagram of size $(N-1) \times K$. As we increase the number $K$, the
diagram $Y''$ starts to cut into the rectangle, see figure \ref{fig3}.
This means that we have to remove a few boxes from those columns and
move them to one of the previous columns. But since all these have
maximal length, every time we take out one of the boxes and move it
to a full column to the left, we lose an entire column with $N$ boxes.
It is easy to see that in total we need to move $n'-r''$  boxes
which make us lose $N(n'- r'')$ boxes altogether. Hence we need
$K=r'' + n'-r'' = n'$ fermions to begin with. Of course,
this number is also sufficient since we can build $Y'$ from the $n'$
fundamentals~$\Box$. This concludes the graphical proof.

Let us now go through a somewhat more formal argument. There is an explicit decomposition \cite{Benkart} found by studying so-called walled Brauer algebras \cite{Turaev,Koike}  which provides us with the irreducible content of the $K$-th tensor power of the SU$(N)$ adjoint representation, at least up to $2K \leq N$,
\begin{align}
\textbf{adj}^{\otimes K}=\sum_{n'=0}^K b^{(K)}_{n'} \sum_{Y',Y'' \vdash n'}
\frac{n'!}{\prod_{(l',m')\in Y'} h\,(l',m')}\frac{n'!}{\prod_{(l'',m'')\in Y''} h\,(l'',m'')}\cdot a\,(Y',Y'') \ .
\label{adj power}
\end{align}
Here we use the notation $Y',Y'' \vdash n'$ to express that both $Y'$ and $Y''$ are partitions of $n'$, the products run over all boxes in $Y'$and $Y''$ and the integers
$h\,(l,m)$ denote the length of a hook that is associated to the box $(l,m)$, see
figure \ref{hook}. A formal definition can be found in appendix A. The multiplicities $b^{(K)}_{n'}$ are of combinatorial nature and explicitly given by
\begin{align}
b^{(K)}_{n'} :=\, \sum_{i=0}^{K-n'} (-1)^{i+K+n'}\,\,i!\,\,\binom{K}{n'}\, \binom{K-n'}{i}\,\binom{i+n'}{i} \ .
\end{align}
The derivation of formula \eqref{adj power} is discussed in more detail in the
Appendix A. What is most important for us right now is that a representation
$a$ composed out of two small Young diagrams of $n'$ boxes can appear on the
right hand side of eq.\ (\ref{adj power}) only when we start with the product
of \mbox{$K=n'$} adjoints  on the left hand side of eq.\ (\ref{adj power}). This
concludes the proof of proposition \ref{prop3}.
\begin{figure}
\begin{center}
\includegraphics[scale=0.2]{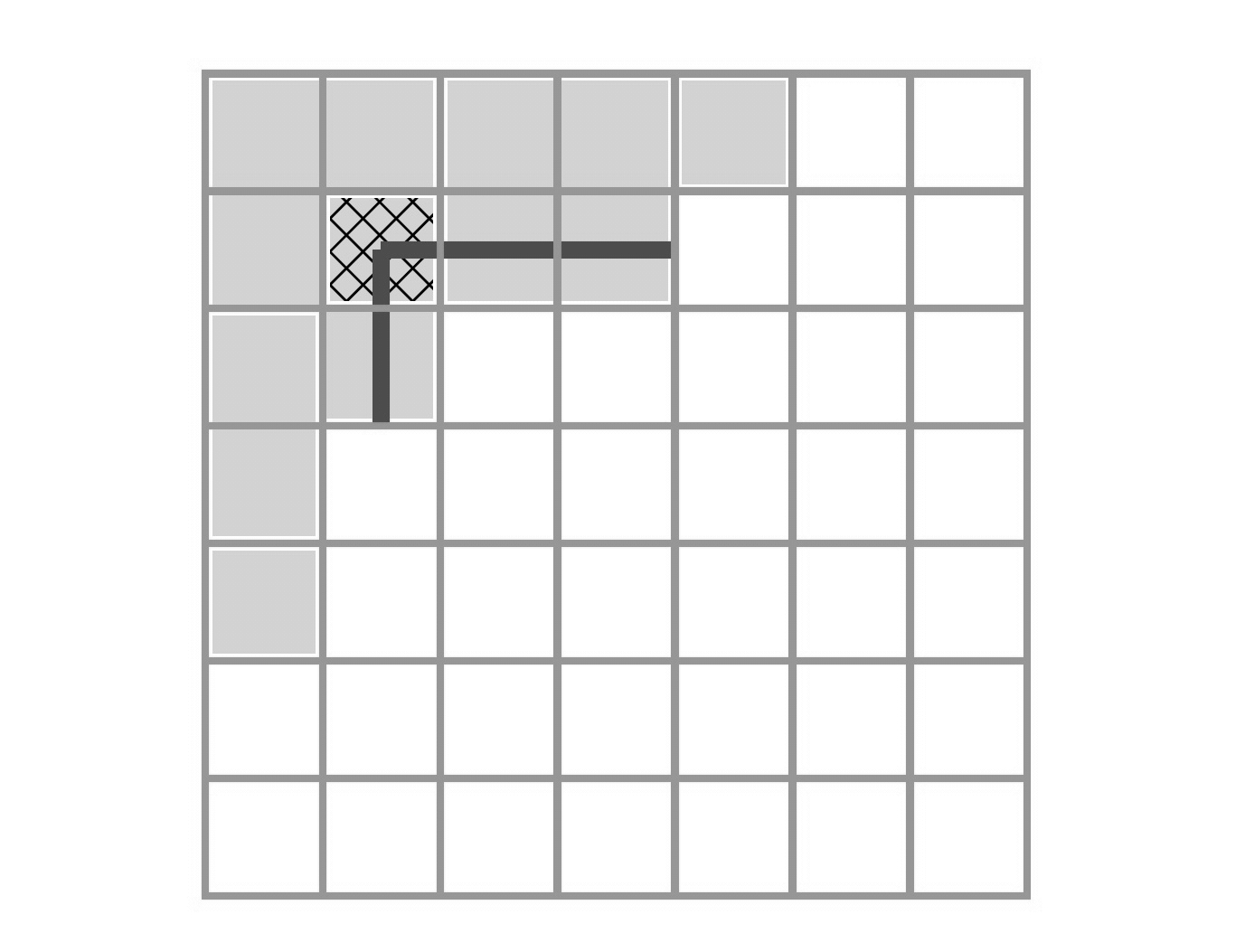}
\caption{For each box $(l,m)$ in the Young diagram $Y$, one can draw a
hook. In the figure we have shaded the box and indicated the hook by the
thick line. The length of the hook, i.e.\ the length of the thick line,
is denoted by $h(l,m)$.}\label{hook}
\end{center}
\end{figure}

Now let us combine the previous three statements. According to proposition
\ref{prop2}, the sectors that can contribute a chiral primary which has finite
weight in the large $N$ limit have $C_2(a) \sim n' N$. Hence, the two
inequalities in proposition \ref{prop1} become
\begin{eqnarray}
 h_\phi & \geq & \frac{K_\phi}{2} - \frac{n'}{3} = \frac{n'}{6}
       + \frac{K_\phi-n'}{2} \ , \\[2mm]
 |Q_\phi| &\leq & \frac{K_\phi}{6} = \frac{n'}{6} + \frac{K_\phi-n'}{6} \ .
\end{eqnarray}
In the second step we have sightly rewritten the bounds. From
proposition \ref{prop3} we know that the quantity $K_\phi - n'$ is
non-negative. Hence, the equality $h_{\phi_\text{cp}} = Q_{\phi_\text{cp}}$
between the conformal weight and U(1) charge of a chiral primary
$\phi_{\text{cp}}$ can only be satisfied for $K_{\phi_\text{cp}} = n'$.
This implies that both the weight and the  U(1) charge of such chiral
primaries,
\begin{equation} \label{fincpresult}
h_{\phi_\text{cp}} = \frac{n'}{6} \quad \mbox{ and } \quad
Q_{\phi_\text{cp}} = \frac{n'}{6} \ ,
\end{equation}
saturate the bounds given in proposition \ref{prop1}. As we explained in the text
below proposition \ref{prop1}, this implies that the state $\phi_\text{cp}$ is
constructed from the fermionic fields $\psi^1_\nu$ only without any derivatives
and components of $\psi^2$. States with these features must transform in the
anti-symmetric tensor power of the adjoint representation. It is actually
possible to work out the precise content of the anti-symmetrized part of the
$k$-th power of adjoint representation for values of $k \leq N-1$,
\begin{align}
\left\{\textbf{adj}^{\otimes k}\right\}_{\text{antisymm}}=\sum_{n'=1}^k
d^{(k)}_{n'} \sum_{Y' \vdash n'} a\,(Y',Y') \ .
\label{adj antipower}
\end{align}
A more detailed discussion and the precise values of the coefficients $d^{(n')}_k$
can be found in Appendix B. What is most important about formula
\eqref{adj antipower}, at least in our present context, is that all representations
that appear in the decomposition are of the form $a(Y',Y'') = a(Y',Y')$. Now we
only need to recall from section~2.3 that such diagonal sectors are associated
with regular chiral primaries to establish our central claim: The chiral primaries
of the large $N$ limit are regular. Let us stress once again that for any given
finite value of $N$, chiral primaries can be constructed that do not satisfy
eqs.\ \eqref{fincpresult} and hence are not regular.

\section{Discussion, conclusion and open problems}

In the preceding section we proved that chiral primary fields in the low energy
limit of multi-color adjoint QCD are regular in the sense we defined in section~2.3. We have seen before that such regular chiral primaries are in one-to-one
correspondence with Young diagrams $Y'$, at least if we approach the multi-color
limit through a sequence of prime numbers $N$. In case $N$ is prime, the only
contribution to the state space \eqref{state_space} that is not simply a diagonal
product of left- and right-movers is the term $\mathscr{H}_\text{fix}$ which
does not contain any regular chiral primaries, see section~2.3. Furthermore,
as we approach the large $N$ theory, the only orbits $\{a\}$ that are
associated with two different diagonal pairs $(Y',Y''=Y')$, namely the
orbits $\{a_\nu\}$, see next to last paragraph in section 2.3, give rise
to regular chiral primaries of weight $h = \nu(N-\nu)/6$. Hence, they are
not part of the spectrum of chiral primaries as $N$ tends to infinity. For
all remaining regular chiral primaries, the orbit is associated with a
unique Young diagram $Y'$. Combining all these facts, we introduce the
symbol $\phi_\text{cp}(Y')$ to denote the unique chiral primary
$$ \phi_\text{cp}(Y') \ \in \ \mathcal{H}_{Y',Y''} \otimes
\overline{\mathcal{H}}_{Y',Y''}   \quad \mbox{ with } \quad
h(\phi_\text{cp}(Y')) = n'= |Y'|\ . $$
As usual in ${\cal N}=(2,2)$ supersymmetric theories, the chiral fields
form a chiral ring which closes under operator product expansions. It is
not difficult to argue
that the chiral ring at large $N$ must be isomorphic
to a standard graded ring of symmetric functions $\Lambda_R=\oplus_{i\in
\mathbb{N}}\, \Lambda^{(i)}_R$,  which is a ring of formal infinite sums
of monomials. Its Hilbert-Poincar\'e series
\begin{align}
\sum_{i\in \mathbb{N}} \text{dim} (\Lambda^{(i)}_R)\,t^i:=\prod_{i=1}^{\infty}\frac{1}{1-t^i} \ ,
\end{align}
i.e.\ the function which generates dimensions of subspaces of grade $i$, is
the generating function of integer partitions. The operator product of two chiral
primaries at large $N$ thus takes the form
\begin{align}
\phi_\text{cp}(Y'_1)\cdot \phi_\text{cp}(Y'_2)\ =\
\sum_{Y'_3}\,\mathcal{C}^{Y'_3}_{Y'_1,Y'_2}\,\phi_\text{cp}(Y'_3),
\end{align}
where $\mathcal{C}^{Y'_3}_{Y'_1,Y'_2}$ are the Littlewood-Richardson coefficients \cite{Littlewood, St-Littlewood}.
The ring is freely generated, e.g.\ by the elementary symmetric polynomials $e_k$,
$k=1,2,\dots$ corresponding to those chiral primaries whose Young diagrams $Y'$
consist of only one column.
Obviously, there is exactly one such generator at each grade. The construction of
generators representing chiral primaries $\phi_\text{cp}(Y')$ corresponding to
other Young diagrams $Y' \neq e_k$ is then performed with the help of the second
Jacobi-Trudi identity.

\begin{figure}
\begin{center}
\includegraphics[scale=0.7]{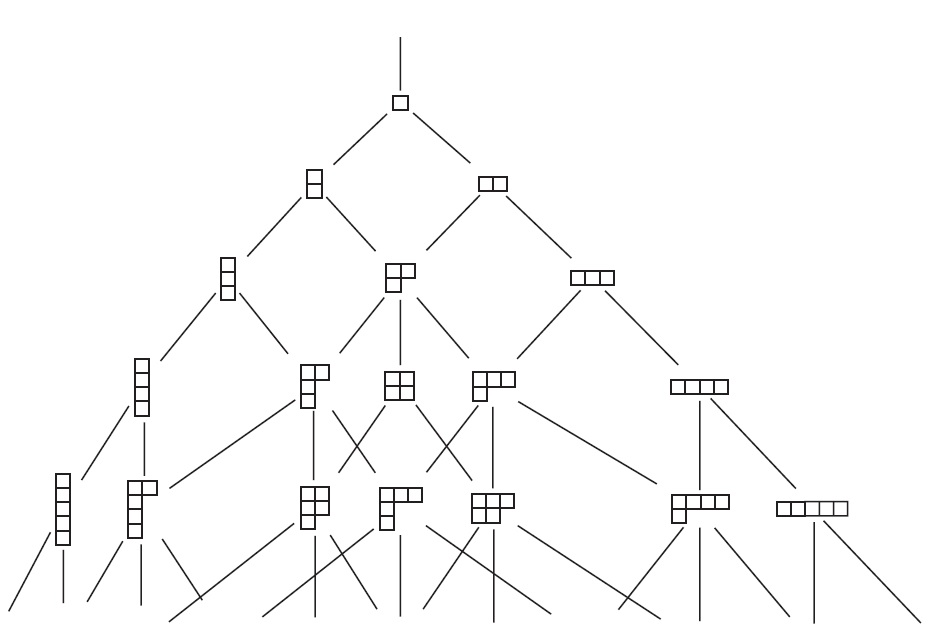}
\caption{Hasse diagram of the Young lattice.
}\label{fig4}
\end{center}
\end{figure}

In the special case of fusion with a chiral primary corresponding to the partition
$f_n$ of one row with $n$ boxes, the Pieri's formula implies
\begin{align}
\phi_\text{cp}(Y'_1)\cdot \phi_\text{cp}(f_n) \ = \
\sum_{Y'_3}\, \phi_\text{cp}(Y'_3),
\end{align}
where the summation goes only over Young diagrams obtained from $Y'_1$ by adding $n$
boxes, no two in the same column. From this formula one can see that an iterative
fusion of the vacuum with the lowest non-trivial chiral primary $\mathcal{C}_{\Box}$
precisely generates the Young lattice (the lattice of Young diagrams ordered by
inclusion). This is a nice way to picture a subring of the chiral ring generated
by the grade $1$ generator alone (see Figure \ref{fig4}).

This concludes our discussion of the results in this paper. Let us recall that
the main motivation of our work stems from the desire to constrain the dual higher
spin or string theory. Let us recall that the family of ``strange metal'' coset
models we analysed in this work has a matrix-like structure in which the central
charge behaves as $c \sim N^2$. While theories with a vector-like dependence $c
\sim N$ have been argued to be dual to Vasiliev higher spin theories in $AdS$,
the dual of strange metal coset theories  with chiral algebra ${\mathcal W}_N$
is believed to possess a much larger symmetry
than Vasiliev theory and could well be a string theory. The background geometry of
such a potential dual string theory is severely constrained by the result we
reported above. Since the spectrum of chiral primaries does not depend on the
string length, we have argued that the background geometry should give rise to
chiral primaries which are in one-to-one correspondence with partitions or Young
diagrams. While we do not have any concrete proposal for now, we want to point out
that infinite families of $AdS_3$ backgrounds with at least ${\cal N} = (2,0)$
supersymmetry have been constructed in \cite{Donos}. It would be interesting to
scan those solutions or apply the methods of Donos et al. in order to find a geometry
that gives rise to the desired chiral ring. Let us note in passing that one chiral
(left-moving) half of the strange metal coset theory was recently argued to arise
in string theory on near horizon geometries of certain fast rotating black holes
in an AdS space \cite{Berkooz}. The constructions of Berkooz et al. provide the
entire state space of the chiral strange metal coset, obviously including all the
chiral primaries we described above. In case the aforementioned  results or methods
do not suffice to identify a dual string background, one might obtain valuable
additional constraints on the dual theory by decomposing the spectrum of the
strange metal coset theory into representations of higher spin symmetries, much
along the lines of \cite{HS-strings}. We plan to come back to these issues
in future research.
\bigskip

\noindent
{\bf Acknowledgements:} It is a pleasure to thank Andrei Babichenko, Micha Berkooz, 
Alessandra Cagnazzo, Rajesh Gopakumar, Tigran Kalaydzhyan, Andrey Kormilitzin, Carlo 
Meneghelli,  Vladimir Mitev, Prithvi Narayan, Vladimir Narovlansky, Kareljan Schoutens 
and Amir Zait for interesting discussions. MI is grateful to the members of the String 
Theory group at the Weizmann Institute for their hospitality during the early stages 
of this work. Some computations were performed on DESY’s IT High Performance Cluster 
(IT-HPC). The research leading to the described results was supported in part by the 
GRK 1670 ''Mathematics Inspired by Quantum Field and String Theory'' of the German 
Science Foundation DFG, by the German-Israeli Foundation under grant number 
I-1-038-47.7/2009 and by the People Programme (Marie Curie Actions) of the European 
Union's Seventh Framework Programme FP7/2007-2013/ under REA Grant Agreement 
No 317089 (GATIS).

\bigskip\bigskip\bigskip

\newpage
\section*{Appendix}
\appendix

\setcounter{equation}{0}
%%%%%%%%%%%%%%%%%%%%%%%%%%%%%%%%%%%%%%%%
\section{Tensor powers of the SU$\mathbf{(N)}$ adjoint representation} \label{appA}
%%%%%%%%%%%%%%%%%%%%%%%%%%%%%%%%%%%%%%%%
In this appendix we discuss the phylogeny of the formula \eqref{adj power} and we
briefly outline its derivation.

Let us start with the following very well known formula for decomposing the $k$-th tensor power of the SU$(N)$ fundamental representation into irreducible SU$(N)$ representations
\begin{align}
\Box^{\otimes k}=\sum_{Y \vdash k}\frac{k!}{\prod_{(i,j)\in Y} h\,(i,j)}\, a\,(Y) \ .
\label{fund power_A}
\end{align}
Here $Y \vdash k$ denotes a Young diagram $Y=(l_1,\dots l_r)$ which has $k$ boxes,
i.e.\ is a partition of $k$, we use $a\,(Y)$ to label the corresponding SU$(N)$ representation and
$$h\,(i,j):=l^{\text T}_j-i+l_i-j+1$$
is defined as the length of a hook $(i,j)$ belonging to the given partition
$Y=(l_1,\dots l_r)$. The product runs over the boxes of the Young diagram $Y$.

The raison d'\^etre of formula \eqref{fund power_A} is the renowned Schur-Weyl duality \cite{Weyl}: The image of the action of the symmetric group $S_k$ on the $k$-th tensor power of the GL$_{N}(\mathbb{C})$ fundamental representation space can be identified with the centralizer algebra of GL$_{N}(\mathbb{C})$ and vice versa. It means that under the joint action of $S_k$ and GL$_{N}(\mathbb{C})$, the tensor power decomposes into a direct sum of tensor products of irreducible modules for these two groups thus yielding formula \eqref{fund power_A}. The coefficient $$\frac{k!}{\prod_{i,j)\in Y} h\,(i,j)}$$ is just the dimension of a corresponding representation of the symmetric group $S_k$.

It turns out that for a tensor power of the adjoint representation $\textbf{adj}^k=\Box^k \otimes \overline \Box^k$ a similar correspondence holds, only that now the symmetric
group algebra gets replaced by a more sophisticated structure known as the {\it walled Brauer algebra} \cite{Turaev,Koike}. The associated decomposition reads
\begin{align}
\textbf{adj}^{\otimes k}=\sum_{m=0}^k b^{(k)}_m \sum_{Y',Y'' \vdash m}
\frac{m!}{\prod_{(i',j')\in Y'} h\,(i',j')}\frac{m!}{\prod_{(i'',j'')\in Y''} h\,(i'',j'')}\cdot a\,(Y',Y'') \ .
\label{adj power_A}
\end{align}
Here $a\,(Y',Y'')$ denotes an SU$(N)$ representation generated from two Young diagrams $Y'$ and $Y''$ according to (\ref{generation}), $Y',Y'' \vdash m$ means that $Y'$ and $Y''$ are Young diagrams corresponding to partitions of $m$. The products in (\ref{adj power_A}) run over boxes of the Young diagrams $Y'$ and $Y''$.
The range of validity here is $k \leq \left\lfloor \frac{N}{2} \right\rfloor$, otherwise not all of the listed representations $a\,(Y',Y'')$ are allowed to appear on the right hand side which results in a reshuffling of the remaining multiplicities. The multiplicities
\begin{align}
b^{(k)}_m :=\, \sum_{i=0}^{k-m} (-1)^{i+k+m}\,\,i!\,\,\binom{k}{m}\, \binom{k-m}{i}\,\binom{i+m}{i}\equiv  \frac{k!}{m!} \sum_{i=0}^{k-m}\frac{ (-1)^{i}}{i!}\, \binom{k-i}{m}\
\label{b^k_m}
\end{align}
are actually the most interesting feature of formula \eqref{adj power_A}. They reflect
the fact that the new algebra replacing the symmetric group algebra in this case is not just a direct product of two copies of the latter. We refer the reader to \cite{Benkart} for background on walled Brauer algebras as well as the representation-theoretic discussion of decomposition formulas, such as the one displayed above.

There is a simple way to argue that the coefficients $b^{(k)}_m$ should have the form \eqref{b^k_m}. Indeed, let us notice that they can be rewritten as
\begin{align}
b^{(k)}_m=(-1)^{k+m}\,\,\binom{k}{m}\,\, _2F_0(m+1,-(k-m);|\,1)\,
\end{align}
where $_2 F_0$ is the hypergeometric function of type $(2,0)$.
It is now straightforward to see that the coefficients $b^{(k)}_m$ actually satisfy the recursion relation
\begin{align}
b^{(k)}_m=\frac{k(k-1)}{k-m}(b^{(k-2)}_m+b^{(k-1)}_m)
\label{recursion}
\end{align}
with the initial conditions $b^{(m-1)}_{m}=0$, $b^{(m)}_m=1$.
This nice recursion readily suggests a way to proceed in proving the decomposition \eqref{adj power_A} with coefficients \eqref{b^k_m}. Acting by induction, the inductive step is just to apply the Littlewood-Richardson rule \cite{Littlewood,St-Littlewood} for multiplying all the Young diagrams present in the decomposition of the $(k-2)$-nd adjoint power by another two adjoint representations, carefully factoring out two hook multipliers which describe adding boxes to 'small' Young diagrams $Y'$ and $Y''$. The relation \eqref{recursion} then allows to disentangle the obtained expression bringing it to the needed $k$-th step's outfit. For the original combinatorial proof involving a generalization of the Schensted insertion algorithm, see \cite{Stembridge}.

\setcounter{equation}{0}
%%%%%%%%%%%%%%%%%%%%%%%%%%%%%%%%%%%%%%%%%%%%%%%%%%%%%%%
\section{Antisymmetric part of the powers of adjoints} \label{appB}
%%%%%%%%%%%%%%%%%%%%%%%%%%%%%%%%%%%%%%%%%%%%%%%%%%%%%%%%%%%%%%%%%%%%%%%%%

Before we begin with our discussion of eq.\ (\ref{adj antipower}) let us
give the precise statement and introduce a bit of additional notation.
According to eq.\ \eqref{adj antipower}, the part of the decomposition
of the SU$(N)$ adjoint power which transforms in the totally antisymmetric
representation of the permutation group $S_k$ is given by
\begin{align}
\left\{\textbf{adj}^{\otimes k}\right\}_{\text{antisymm}}=\sum_{m=1}^k
d^{(k)}_m \sum_{Y' \vdash m} a\,(Y',Y') \ .
\label{adj antipower_B}
\end{align}
Here $a\,(Y',Y')$ denotes an SU$(N)$ representation generated from two
Young diagrams $Y'=Y''$ according to (\ref{generation}), $Y' \vdash m$
means that $Y'$ is a Young diagram satisfying $|Y'|=m$, i.e.\ is a
partition of $m$.

The coefficients $d^{(k)}_m$ read
\begin{align}
d^{(k)}_m :=\sum_{1 \leq m_1 < \dots < m_{k-1}\leq m-1}
r_{m_1}
r_{m_2-m_1}\dots r_{m_{k-1}-m_{k-2}} r_{m-m_{k-1}}
\end{align}
where $r_m$ are expressed as
\begin{align}
r_m :=\frac{1}{m!}\,
\left(\frac{d}{dq}\right)^{m-1}\left[\left(\frac{q}
{\frac{\phi^3(q^2)}{\phi(q)\phi(q^4)}-1}\right)^m\right]_{q=0}\equiv \frac{1}{m!}\, \left(\frac{d}{dq}\right)^{m-1}
\left[\left(\frac{q}{\phi(-q)-1}\right)^m\right]_{q=0}
\end{align}
and $\phi$ is the Euler function
\begin{align}
\phi(q):=\prod_{i=1}^{\infty} (1-q^i).
\end{align}
The range of validity of the formula (\ref{adj antipower_B}) is restricted by $k
\leq N-1$, otherwise not all of the listed representations $a\,(Y',Y')$ are allowed
to appear on the right hand side which results in a reshuffling of the remaining
multiplicities.

We checked this formula by direct computation up to $k=9$. Unfortunately, we were not able to find it in the literature. One immediate aspect to notice is that upon applying the Lagrange inversion formula, the coefficients $d^{(k)}_m$ turn out to be just coefficients of the series expansion of $q^k(\phi)$, where $q(\phi)$ denotes the function inverse to $\phi(-q)-1$ around $q=0$.

The parts of the SU$(N)$ adjoint powers' decomposition transforming in other representations of the symmetric group will, of course, involve the 'non-diagonal' representations $a\,(Y',Y'')$, with $Y'\neq Y''$, to yield the full decomposition \eqref{fund power_A} when summed up over all representations of the symmetric group $S_k$. It is tempting to speculate that the multiplicities in those partial decompositions may be characterized by other modular forms replacing the Dedekind eta $\eta(q)=q^{\frac{1}{24}}\phi(q)$. The exact formulae of this type will be discussed elsewhere.

%%%%%%%%%%%%%%%%%%%%%%%%%%%%%%%%%%%%%%%%%%%%%%%%%%%%%%%%%%

\end{document}